\documentclass[11pt,twoside]{article}
\usepackage{asp2004}
\usepackage{epsf}
\usepackage{lscape}

\begin{document}
\title{Light Variations of the Anomalous Central Star of Planetary
Nebula Sh~2-71}

\author{Z. Mikul\'a\v sek$^{1,\,3}$, A. Skopal$^{2}$, M.
Zejda$^{1,\,3}$, O. Pejcha$^{3}$, L. Kohoutek$^{4}$, D.~Motl$^{3}$,
A.\,A. Vittone$^{5}$, L. Errico$^{5}$}

\affil{$^1$Institute of Theoretical Physics and Astrophysics,
Masaryk University, Kotl\'a\v{r}sk\'a 2, CZ-611\,37 Brno, Czech
Republic, mikulas@ics.muni.cz}

\affil{$^2$Astronomical Institute of the Slovak Academy of Science,
SK-059\,60 Tatransk\'a Lomnica, Slovak Republic}

\affil{$^3$N. Copernicus Observatory and planetarium, Krav\'i hora
 2, CZ-616\,00 Brno, Czech Republic}

\affil{$^4$Hamburg Observatory, D-21029 Hamburg, Germany}

\affil{$^5$INAF Osservatorio Astronomico di Capodimonte, I-80131
Napoli, Italy}

\begin{abstract}
We present an analysis of light variations in
\emph{UBV}$\!(\!RI)_{\rm{C}}$ of the anomalous object in the center
of planetary nebula Sh~2-71. We refined the linear ephemeris of the
light maxima to $\mathrm{JD_{max}}=2449862.0+68.101~(E-96)$, but
also identified long-term, obviously non-periodic variations. The
latter manifest themselves in large O-C shifts, a variable profile
of light curves (hereafter LC) and changes in the mean brightness of
the object. Our spectroscopic observations suggested the presence of
a superdense nebula in the center of Sh~2-71.
\end{abstract}

\section{Introduction}

The variability of the central object in the planetary nebula
Sh~2-71 was discovered by Kohoutek \cite{koh}. Jurcsik \cite{jur93}
revealed a 68.064-day period in the nearly parallel
\emph{UBV}$\!(\!RI)_{\rm{C}}$ light variations of the central star.
Mikul\'a\v{s}ek et al. \cite{mik} enlarged photometric data by their
\emph{V}$\!(\!RI)_{\rm{C}}$ observations from 1999--2002 and using
all \emph{V} data improved the period to 68.132 days. They noticed
differences between LCs of various colours and a scatter in the
(O-C)-diagram.

\section{Observations}

This study presents first results of PCA analysis of light behaviour
of the object based on all available photometric data obtained
during the time interval of 29 years. We have processed a large set
of 4268 observations from three sources:

\begin{center}
\begin{tabular}{|l|r|rrrrr|r|}
\hline
Author  & years  & $U~$ &  $B~$  &  $V~$  &  $R_{\mathrm{C}}$ & $I_{\mathrm{C}}$ & sum\\
\hline
Kohoutek & 77--79 & 81 & 82 & 100& -~~ & -~~ & 263 \\
Jurcsik & 90--93 & \,392 & \,401 & \,403 & \,403 &\,402& \,2001 \\
Our paper\ \, & 99--05 & -~~ &  -~~ &  750 &794& 460& 2004\\
\hline
 &        $N_{\mathrm{tot}~}$ & 473 & 483& 1253& 1197& 862& ~{\bf 4268}\\
 \hline
\end{tabular}
\end{center}
\begin{figure}[p!th]
\begin{center}
\resizebox{0.39\hsize}{!}{\includegraphics{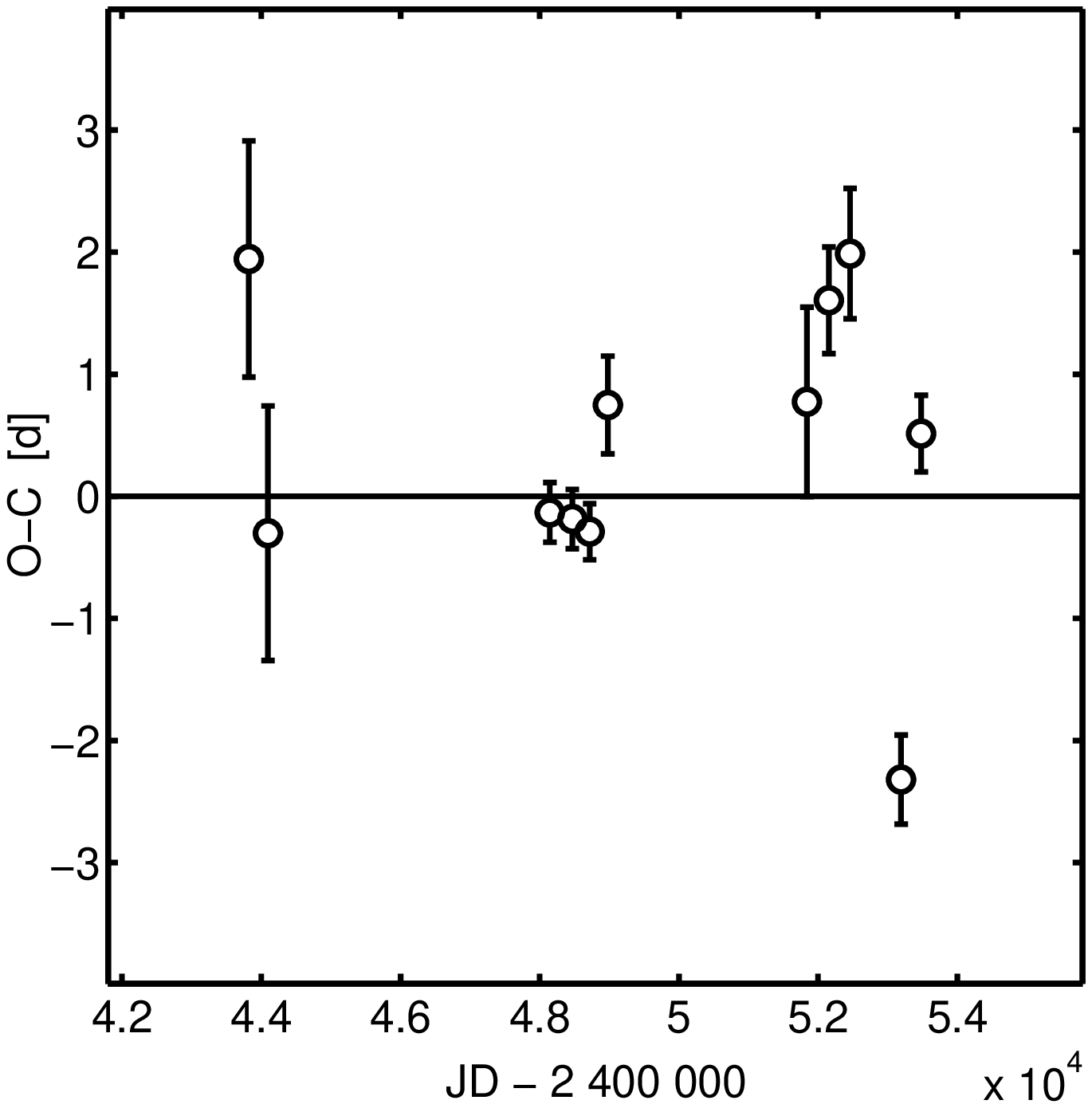}}
\resizebox{0.463\hsize}{!}{\includegraphics{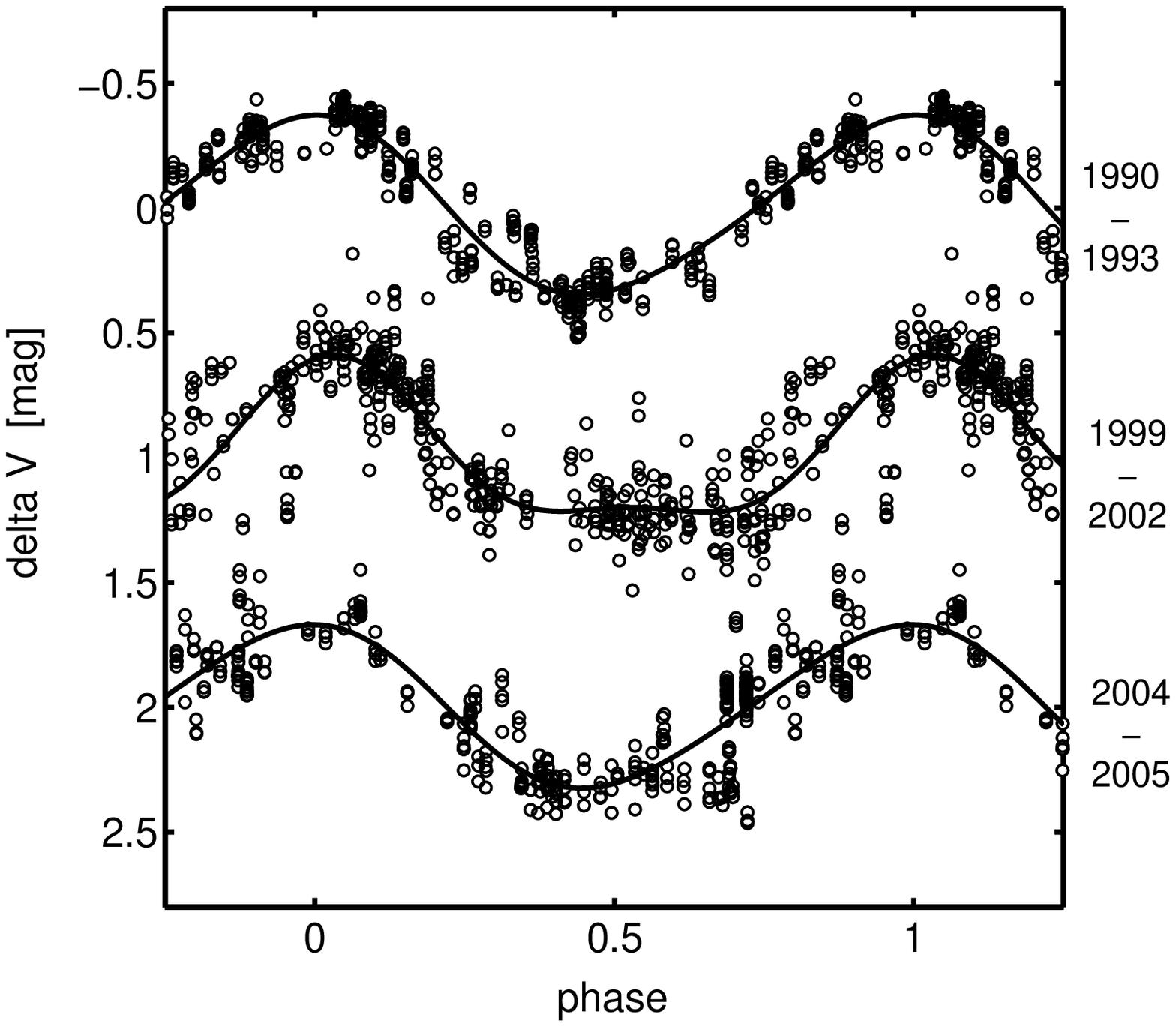}}
\end{center}
\caption{The O-C diagram and the long-term variations in the
\emph{V}-LC.}
\end{figure}
\noindent 1. Partly unpublished \emph{UBV} observations made by
Kohoutek at Wise Observatory at Mitzpe Ramon, Israel, in 1977--79,
at ESO at La Silla, Chile and at the Hamburg-Bergedorf Observatory,
Germany, in 1979.

\noindent 2. Unpublished \emph{UBV}$\!(\!RI)_{\rm{C}}$ observations
made by Jurcsik \cite{jur03} at Konkoly Observatory in 1990--93.

\noindent 3. Unpublished \emph{V}$\!(\!RI)_{\rm{C}}$ CCD photometry
made by M. Zejda, P. H\'{a}jek, O. Pejcha, J. \v{S}af\'{a}\v{r}, P.
Sobotka, D. Motl and others at Brno (1999--2005) and Vy\v{s}kov
(2001--2005) Observatories.

\section{Periodic light variations}

Analysis of photometric data confirmed the basic 68-day periodic
variation of the object in all passbands. Application of own robust
(Mikul\'a\v{s}ek et al. 2003) PCA code {\small \sf PERISH} to all
4268 individual photometric observations allowed us to derive the
linear ephemeris for timings of the light maxima as

\begin{equation}
\mathrm{JD_{max}}=(2\,449\,862.02\pm0.32)\!+\!(68.101\pm0.010)(E\!-\!96).
\end{equation}

The profile of the \emph{U}-LC differs from those observed in other
passbands.

\section{Long-term variations}

Photometric monitoring of Sh~2-71 during the last 15 years
(1990--2005) revealed long-term variations in its light. Residuals
resulting from the linear ephemeris (1) suggest a non-periodic
shifts of LC as whole (Fig.~1a). Further we selected the \emph{V}-LC
during three periods (1990--93, 1999--2002, 2004--05) well covered
by observations (Fig.~1b). During the 1999--02 period the minimum
became flat from the phase 0.3 to 0.75.
In addition, a different average \emph{V}-magnitudes were measured
during 1990-93, 1999-2002 and 2004-05. It was 13.48, 13.59 and
13.50, respectively.
\begin{figure}[p!th]
\begin{center}
\resizebox{0.83\hsize}{!}{\includegraphics{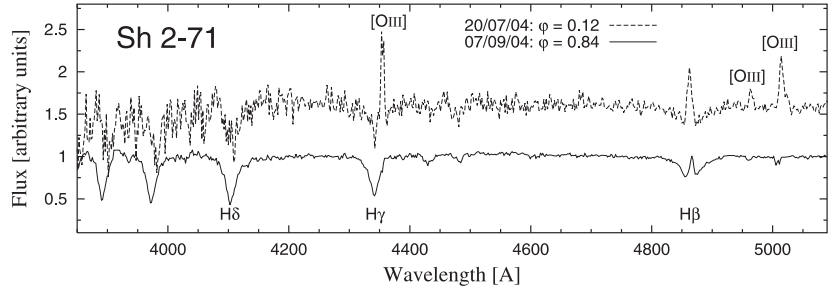}}
\end{center}
\caption{Low-resolution spectra of the core of Sh~2-71.}
\end{figure}

According to the binary model of the core of Sh 2-71 (Cuesta \&
Phillips, 1993) we interpret the periodic variation in the LC as due
to the orbital motion of the binary components. The variability
could be caused by different projection of an optically thick inner
nebula due to the orbital motion. The long-term changes may be
attributed to a variable mass transfer between the components.

Our two low-resolution spectra of the central object of Sh~2-71
carried out at the Loiano Observatory with the BFOSC spectrograph on
20/07/04 (phase 0.12) and 07/09/04 (phase 0.84) support this
interpretation. They indicate strong change in the ionization
conditions at these phases (Fig.~2). The ratio
$R=(F(4958)+F(5007))/F(4363)=3.3$ suggests a superdense inner [O
III] nebula likewise in symbiotic binaries \cite{sk+01}.

\section{Conclusions}

We refined the average linear ephemeris of the periodic variation in
all \emph{UBV}$\!(\!RI)_{\rm{C}}$ LCs. Revealed long-term variations
manifest themselves in found changes of (i)~timing the light maxima
as suggested by the O-C residuals, (ii)~the profile of the
\emph{V}-LC, which occasionally becomes flat for a large phase
interval, and (iii)~the variable average brightness of the nucleus.

Our low-resolution spectroscopy revealed apparent differences
between the spectra taken in different phases. Fluxes of [O III]
nebular lines suggest a very dense inner nebula of Sh~2-71.

\acknowledgements We thank J. Jurcsik for providing us with her
data. This work was granted by GA\,\v{C}R 205/04/1267, 205/04/2063,
MVTS SR-\v{C}R 128/04 and SAS No.~2/4014/4.

\end{document}